\documentclass[conference]{IEEEtran}
\IEEEoverridecommandlockouts

\usepackage{url}            
\usepackage{multirow}
\usepackage{subcaption}

\usepackage{hyperref}
\usepackage{url}
\usepackage[mathletters]{ucs} 
\usepackage[utf8x]{inputenc}
\usepackage{pifont}
\usepackage{multirow}
\usepackage{comment}
\usepackage{amsmath}
\usepackage[T1]{fontenc}    
\usepackage{hyperref}       
\usepackage{url}            
\usepackage{booktabs}       
\usepackage{amsfonts}       
\usepackage{microtype}      
\usepackage{todonotes}
\usepackage{graphicx}
\usepackage{wrapfig}
\usepackage{xcolor}
\usepackage{enumitem}
\usepackage{algorithm2e}

\usepackage{listings}
\usepackage{lstautogobble}

\def\BibTeX{{\rm B\kern-.05em{\sc i\kern-.025em b}\kern-.08em
    T\kern-.1667em\lower.7ex\hbox{E}\kern-.125emX}}
\begin{document}

\title{Code Compliance Assessment as a Learning Problem}

\author{\IEEEauthorblockN{Neela Sawant}
\IEEEauthorblockA{\textit{Amazon} \\
Bengaluru, India \\
nsawant@amazon.com}
\and
\IEEEauthorblockN{Srinivasan H. Sengamedu}
\IEEEauthorblockA{\textit{Amazon} \\
Seattle, USA \\
sengamed@amazon.com}
}

\maketitle

\begin{abstract}
Manual code reviews and static code analyzers are the traditional mechanisms to verify if source code complies with coding policies. However, these mechanisms are hard to scale. We formulate \emph{code compliance assessment} as a machine learning (ML) problem, to take as input a natural language policy and code, and generate a prediction on the code's compliance, non-compliance, or irrelevance. This can help scale compliance \emph{classification} and \emph{search} for policies not covered by traditional mechanisms.

We explore key research questions on ML model formulation, training data, and evaluation setup. The core idea is to obtain a joint  code-text embedding space which preserves compliance relationships via the vector distance of code and policy embeddings. As there is no task-specific data, we re-interpret and filter commonly available software datasets with additional pre-training and pre-finetuning tasks that reduce the semantic gap. 

We benchmarked our approach on two listings of coding policies (\emph{CWE} and \emph{CBP}). This is a zero-shot evaluation as none of the policies occur in the training set. On \emph{CWE} and \emph{CBP} respectively, our tool \emph{Policy2Code} achieves classification accuracies of (59\%, 71\%) and search MRR of (0.05, 0.21) compared to CodeBERT with classification accuracies of (37\%, 54\%) and MRR of (0.02, 0.02). In a user study, 24\% Policy2Code detections were accepted compared to 7\% for CodeBERT. 
\end{abstract}


\section{Introduction}

Coding policies are natural language rules or best practices on diverse topics like security, exception handling, and input validation. Code compliance assessment, e.g., verifying that source code is compliant  with coding policies is an important but non-trivial problem. Consider two policies from the well-known \emph{Common Weakness Enumeration} (CWE) listing \cite{cweSite}. 

\begin{enumerate} 
\item CWE-396\footnote{\url{https://cwe.mitre.org/data/definitions/396.html}} - \textit{``Catching overly broad exceptions promotes complex error handling code that is more likely to contain security vulnerabilities''}. Now consider this code snippet:
\begin{lstlisting}[basicstyle=\scriptsize,language=Java,frame=single,breaklines=true,autogobble]
private static byte[] decrypt(byte[] src, byte[] key) throws RuntimeException {
    try {
        SecureRandom sr = new SecureRandom();
        DESKeySpec dks = new DESKeySpec(key);
        ...
        Cipher cipher = Cipher.getInstance(DES);
        cipher.init(Cipher.DECRYPT_MODE, securekey, sr);
        return cipher.doFinal(src);
    } catch (Exception e) {
        throw new RuntimeException(e);
    }
}
\end{lstlisting}
This code  is non-compliant with CWE-396 as the \textit{`catch (Exception e)'} statement broadly covers all exception types. Instead, specific exceptions such as \textit{NoSuchAlgorithmException} or \textit{NullPointerException} should be used. 

\item CWE-197\footnote{\url{https://cwe.mitre.org/data/definitions/197.html}} - \textit{``Truncation errors occur when a primitive is cast to a primitive of a smaller size and data is lost in the conversion''}. Now consider this code snippet: 
\begin{lstlisting}[basicstyle=\scriptsize,language=Java,frame=single,breaklines=true,autogobble]
public static byte[] toByteArray(InputStream input, long size) throws IOException {
    if (size > Integer.MAX_VALUE) {
        throw new IllegalArgumentException(""Size greater than Integer max value: "" + size);
    }
    return toByteArray(input, (int) size);
}
\end{lstlisting}
This code is compliant with CWE-197  because the casting operation  \emph{`(int) size'} maps variable \emph{size} from primitive \emph{long} to another primitive \emph{int} of smaller size, and this operation is protected against truncation errors via the preceding \emph{if} comparison on integer maximum value.
\end{enumerate} 
 
Code compliance assessment requires a deep understanding of program behavior and associated policies. This task is subsequently carried out via manual code reviews and static code analysis. However, both approaches are hard to scale to increasing and evolving coding policies across diverse programming languages and frameworks. The global market size for static code analyzers is projected to reach US \$2002 million by 2027, from US \$748.1 million in 2020, at a cumulative annual growth rate of 15.1\% during 2021-2027~\cite{marketResearch}. Increased automation can help with this goal. 

Our objective is to explore how machine learning can help scale code compliance assessment - \emph{``Can we automatically label a code as compliant or non-compliant (or irrelevant) given any natural language policy input?''} This has two use-cases: (a) \emph{classification} - detecting non-compliant codes for policies not covered by manual reviews and static analyzer rules, and (b) \emph{search} - finding compliant and non-compliant code examples that will serve as test cases in developing new static analysis rules. While there is prior work in detecting buggy codes \cite{habib2019neural, pradel2018deepbugs, neuralBugWithContextCodeRep, 9252010} as well as natural language code search \cite{sachdev2018ncs, cambronero2019WDLMCSmisc, gu2018DCS}, no solutions exist today that can simultaneously determine code relevance as well as compliance and non-compliance. In this paper, we set up a learning framework and explore three research questions. 
\begin{itemize}
\item \emph{Learning objective} - Compliance assessment is similar, but more fine-grained than relevance assessment. Both the compliant and non-compliant examples are relevant to a coding policy. \emph{How can we frame the learning problem?} We propose learning different embeddings to represent compliant and non-compliant \textit{facets} of natural language policies. We formulate a representation learning approach where the relationships between the compliant and non-compliant policy facets and compliant, non-compliant, and irrelevant code examples are preserved via the vector distances between their embeddings. Representation learning can scale compliance assessment in a \emph{zero-shot} setting \cite{zeroshot} as new policies and code examples can be mapped to their embeddings, even if they are not part of the original training dataset. 
 
\item \emph{Training setup} - There are no task-specific labeled training datasets on coding policies associated with their examples. Curating new training datasets can be prohibitively expensive. \emph{What training data can be used?} We propose repurposing general-purpose code and documentation datasets and propose additional training and filtering tasks to reduce the semantic gap.

\item \emph{Evaluation setup} - \emph{How to quantify the performance of learning-based tools for code compliance assessment?}  We contribute a new benchmark that  reinterprets and repurposes well known listings of coding issues curated by program analysis experts: (a) Common Weaknesses Enumeration (CWE) \cite{cweSite} and (b) Coding Best Practices (CBP) from Amazon CodeGuru test suite \cite{codeguru}. We consider weakness or issue descriptions as policies, buggy examples as non-compliant and correct code examples as compliant examples. For realistic search setting, we further add  27K unlabeled code snippets from the CodeSearchNet test dataset \cite{husain2020codesearchnet}. We use standard metrics to measure performance (\emph{accuracy} for  classification and \emph{mean reciprocal rank (MRR)} for search) to compare against two strong baselines - CodeBERT and a multi-class model.
\end{itemize} 

We develop a new tool \emph{Policy2Code} by exploring policy representations, loss formulation, and training schemes. On CWE and CBP respectively, \emph{Policy2Code} achieves classification accuracies of (59\%, 71\%) and search MRR of (0.05, 0.21). In comparison, CodeBERT achieves classification accuracies of (37\%, 54\%) and MRR of (0.02, 0.02) respectively. The multi-class model achieves classification accuracies of (54\%, 60\%) and MRR of (0.03, 0.13) respectively. We set up Policy2Code search engine to on-demand retrieve compliant and non-compliant examples of new policies.  A user study with program analysis experts showed that 24\% detections from \emph{Policy2Code} were accepted compared to 7\% detections from CodeBERT and 15\% by the multi-class model. \emph{Policy2Code} is better than other ML alternatives, but still not practical as a stand alone code compliance tool. We expect it to augment manual efforts in the near future. We hope our work spawns future research in alternative and robust formulations for automated code compliance assessment.

\section{Related Work}\label{sec:relatedWork}

We discuss machine learning for code applications as well as key ideas in information retrieval and representation learning. 

\subsection{Machine Learning for Code} 
Machine learning and deep learning in particular, have advanced many software engineering tasks such as variable naming \cite{allamanis2018learning}, comment understanding \cite{9610635}, bug detection \cite{habib2019neural}, code review generation \cite{siow2019core},  and code and documentation synthesis \cite{dlForCodeModelGen, 9610724}.  Deep learning research has benefitted from efficient transformer models \cite{attentionIsAllYouNeed} such as CodeBERT \cite{feng2020codebert}  - a 124M parameter text-code encoder with impressive performance on  documentation generation and  code search tasks. Another useful concept is that of pre-training, allowing data-hungry deep learning models to learn from large unlabeled or weakly labeled datasets on related tasks, so that the models can be subsequently fine-tuned using small task-specific training datasets \cite{devlin-etal-2019-bert, ramachandran-etal-2017-unsupervised, muppet}. CodeBERT is trained with CodeSearchNet training dataset for two tasks: masked language modeling and code-text relevance (method body to method descriptions). 

\emph{Neural Bug Finding} is typically modeled in a supervised setting using labeled training examples for a pre-determined set of issues. Training examples may be obtained from existing static analyzer detections \cite{habib2019neural} or synthesized with transformations \cite{pradel2018deepbugs}. Much of the research is devoted to analyzing the effect of richer program representations. For example, Wang et al.~\cite{neuralBugWithContextCodeRep} show that combining different types of contexts (the Program Dependence Graph (PDG),  Data Flow Graph (DFG) and the method under investigation) improves performance of bug detection in pre-determined categories. 

Earlier \emph{Code Search} systems predominantly relied on string matching, and in some cases aiming to improve performance with code structures \cite{linstead2009Sourcerer}, API matching \cite{lv2015codehow},  API usage patterns  \cite{mcmillan2011portfolio, raghothaman2016swim}, and query expansion \cite{lu2015codeWordNet}.  More recent neural code search systems perform search based on the vector distance between their embeddings. This research focuses on three aspects (a) code representations, for example using method names and API sequences \cite{gu2018DCS}, parse trees \cite{allamanis2015bimodal}, and control flow graphs \cite{yao2019mman, ling2020deep}, (b) training strategy, for example, unsupervised  \cite{sachdev2018ncs, 9252051} or supervised \cite{allamanis2015bimodal, feng2020codebert, cambronero2019WDLMCS, martins2020concra}, and (c) loss formulations, for example, contrastive loss or \cite{allamanis2015bimodal, feng2020codebert} or triplet loss  \cite{gu2018DCS, yao2019mman, ling2020deep, martins2020concra, yao2019coacor}.

\subsection{Representation learning} 
To the best of our knowledge, there are no settings for fine-grained compliance assessment of code and policy pair. We review key research in general information retrieval and representation (metric) learning \cite{Kaya2019DeepML, luan2020sparse, facetedSearch, beyondFaceted}. Representation learning has the attractive property that once the embedding space is learned, novel input can be mapped without additional training, essentially a \emph{zero-shot} setting \cite{zeroshot}.

A common ranking mechanism is \emph{triplet loss} that models relationships among three data points - an anchor, a positive example, and a negative example. In traditional code search, they map to natural language query (such as a method description) to relevant example (such as the corresponding method body) and an irrelevant example (such as the body of an unrelated method). \cite{Manmatha2017SamplingMI, SchroffKP15, hermans2017defense}. The algorithm learns to minimize the distance between anchor-positive pairs and maximize the distance between the anchor-negative pairs. Triplet mining is an active area of research \cite{Manmatha2017SamplingMI, SchroffKP15, hermans2017defense}.  There are very few settings involving higher order losses for fine-grained differentiation. For example, Chen et al. \cite{chen2017quadruplet}  introduced a quadruplet loss network for person verification task. A training example in this task consists of four data points $x_i, x_j, x_k, x_l$ where $x_i, x_j$ belong to the same class (e.g., considered the same person) and $x_k$ and $x_l$ belong to two other classes (faces from two other people). 

\subsection{Other domains} 
Our objective may seem similar to the sentiment analysis task  \cite{luo-etal-2019-towards, DBLP:journals/corr/abs-1804-06437} in natural language understanding. The objective of this task is to classify a natural language text into positive, negative, or neutral polarities. The difference is that the polarities of natural language are not comparable with the polarities of coding policies. The semantics of a polarity is globally consistent for natural language. For example, both these sentences - \textit{`food is exceptionally good'} or \textit{`the concert was exceptionally good'}, convey similarly positive emotions about two different entities. On the other hand, the compliance or non-compliance of code is always conditioned on the specific policy under consideration.  The same  snippet may be compliant with one policy but non-compliant with another (for example, adheres to exception handling requirements, but is not secure). 


\section{Learning Problem Formulation}\label{sec:problemFormulation} 
Let $r$ and $c$ denote a natural language policy and a code example. Let $y \in \{+, -\}$ denote the compliant and non-compliant facets respectively, and let irrelevance be denoted by $\sim$. Our objective is to learn a bimodal embedding space for $r$ and $c$ such that the the compliant, non-compliant (and irrelevance) relationships are represented via the vector distances between their embeddings. In contrast to code search, where natural language query has a single representation, we propose to learn separate $r^+$ and $r^-$ embeddings for fine-grained differentiation. For compliant and non-compliant example search, we then operate with two separate query inputs $r^+$ and $r^-$. Embedding of  $r^+$ should be closer to $c^+$, followed by $c^-$, and dissimilar to that of $c^\sim$. For non-compliant search, $r^-$ is closer to $c^-$, followed by $c^+$, and far from $c^\sim$. Such space naturally supports search and classification based on relative distances and learned thresholds. New policies and codes can be mapped to their embeddings and evaluated in zero-shot setting. We now discuss two aspects of representation learning - policy input representation and loss functions. 

\subsection{Facet Policy Input Representation} 
Policy representation is a complex problem. A policy contains a general relevance context and a specific compliance context. For example, reconsider  \textit{(CWE-396): Catching overly broad exceptions promotes complex error handling code that is more likely to contain security vulnerabilities}. This policy is relevant to all code snippets with exception handling. The compliance context refers to using broad exception types (\emph{Exception}) versus more specific exceptions (\emph{NoSuchAlgorithmException} or \emph{NullPointerException}). Policy representations can incorporate domain knowledge and advanced parsing of natural language text. In the context of this paper however, we explore two purely learning based variations inspired from prior work in NLP and Computer Vision. 

\subsubsection{Facet-prefixed Policy}
This idea is similar to T5 query format \cite{raffel2020T5} with remarkable success in NLP applications. Natural language policy description $r$ is prefixed with facet token $y$. The concatenated representation $y:r$ is passed through an encoder to generate faceted policy embedding. We use a transformer-based encoder with self-attention, so that the facet token organically attends to the various policy text components. 

\subsubsection{Facet-masked Policy}
This approach externally combines the facet token with the policy input using a parametric model. We use an implementation of conditional masking~ \cite{veit2017csn}, which aims to learn as many mask vectors as the number of conditions (e.g., facets in our setting). One of the mask vector is selected based on the facet $y$ and multiplied with embedding of policy $r$, essentially projecting the policy into a different subspace conditioned on the facet label. See Appendix \ref{sec:facetLearn} for details.  

\subsection{Loss Functions}
\begin{figure}[ht!]
\centering
\includegraphics[width=\columnwidth]{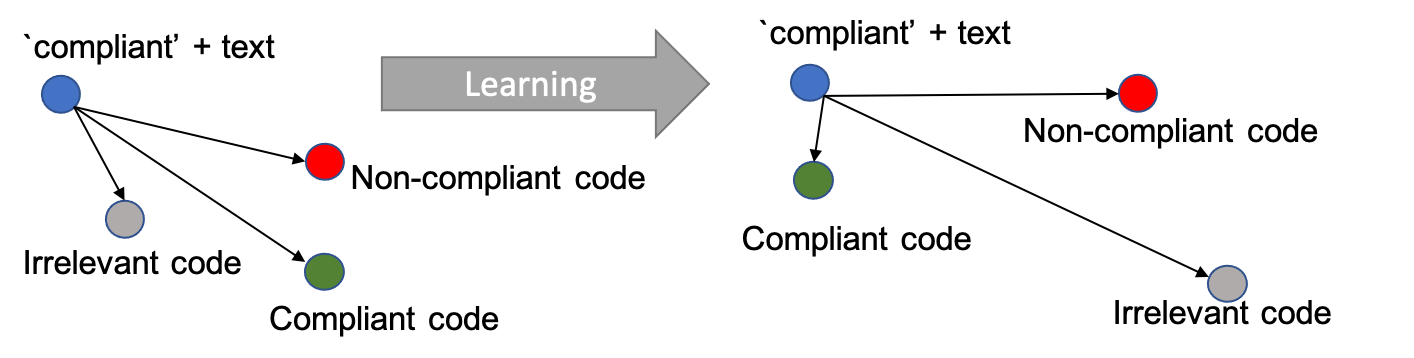}
\caption{Illustration of compliance search task (with compliant facet): The learning objective is to minimize the vector distance between the faceted policy description and a compliant code embedding, and maximize the distance with respect to non-compliant and irrelevant code embeddings.} 
\label{fig:quadrupletIllustration}
\end{figure}

Figure \ref{fig:quadrupletIllustration} illustrates the learning objective using an example of compliance search task. Given a training quadruplet $(r^+, c^+, c^-, c^\sim)$, we want to learn embeddings that minimize the relative distances between matched pairs and maximize the distances between unmatched pairs. We test two formulations.

\subsubsection{Quadruplet Loss (QL) } 
Quadruplet loss is an extension of triplet loss computed as follows: 
\begin{equation}
\begin{aligned}[b]
L^+_{quad} &= \sum_{r^+ ,c^+,c^-}^{N} [d(f_{r^+}, f_{c^+})^2 - d(f_{r^+}, f_{c^-})^2 + \alpha_1]_{+}  \\ &+ \sum_{r^+ ,c^-, c^\sim}^{N} [d(f_{r^+}, f_{c^-})^2 - d(f_{r^+}, f_{c^\sim})^2 + \alpha_2]_{+}
\end{aligned}
\label{eqn:compliantLoss}
\end{equation}
where $f_x$ denotes the embedding of sequence $x$, $d(., .)$ denotes the distance function between two embeddings, and $[]_{+}$ indicates that only positive values contribute towards the loss calculation. $\alpha_2 > \alpha_1 > 0$ are tunable margin parameters for compliance and relevance differentiation, respectively. The loss for the non-compliant facet can be computed similarly over quadruplet $(r^-, c^-, c^+, c^\sim)$. Finally, the total loss is computed as $L_{quad} = \frac{1}{2} (L^+_{quad} + L^-_{quad})$. 

The naive quadruplet loss formulation has two limitations. Eqn.~\ref{eqn:compliantLoss} is anchored around text only. It does not encode constraint for classification task because there is no anchoring around code and no constraint to ensure that embeddings of $r^+$ and $r^-$ are different given a code. Another limitation is from practically available data. In general purpose datasets, very few examples will contain both compliant and non-compliant codes. However, there is an abundance of datasets with partial view (some contain only relevant examples, some with only compliant or non-compliant examples, etc.). To be able to use the latter datasets, we need an alternative formulation as below. 

\subsubsection{Bimodal Multi-task Triplet loss (BMT) } 
In this formulation, the quadruplets $(r^+, c^+, c^-, c^\sim)$ and $(r^-, c^-, c^+, c^\sim)$ are un-pivoted into a batch $B$ of labeled examples $(x, l)$, where $x$ can be either a text or code (e.g., $x$ is bimodal) and label $l$ is generated on the fly such that only the $x$ originating from paired policy and facets share that label $l$. For example, the quadruplets mentioned above can be un-pivoted into a list $[(r^+, l^1), (c^+, l^1), (r^-, l^2), (c^-, l^2), (c^\sim, l^\sim), \ldots]$. Each irrelevant example $c^\sim$ is assigned a unique label $l^\sim$ to prevent anchor-positive matching with another entity. We then calculate a \emph{bimodal multi-task triplet (BMT)}  loss over possible valid triplets $(a, p, n)$ where $a$, $p$, and $n$ are anchor, positive, and negative sequences in the batch such that $a$ and $p$ share the same label and $n$ has a different label. The term \emph{multi-task} denotes the simultaneous modeling of different class labels.  \begin{equation}
L_B  =  \sum_{(a, p, n); l(a)=l(p), l(a) \ne l(n)}  [d(f_a, f_p)^2 - d(f_a, f_n)^2 + \alpha]_{+}
\label{eqn:multiTaskTriplet}
\end{equation}
where $\alpha$ is a common margin which acts as a threshold to separate the irrelevant examples from the relevant ones. The formulation naturally allows both code and text entities to serve as anchors. Datasets with partial labels (only relevant, compliant, and/or non-compliant)  can be incorporated into training. Compliance search task is served by selecting codes based on their distances from faceted policy representations. For compliance classification, we use the threshold $\alpha$ to determine if a code $c$ is relevant or irrelevant with respect to the average policy embedding $(r^+ + r^-)/2$. If it is deemed relevant, we use the distances with respect to $r^+$ and $r^-$ to determine the closest facet as compliance label. The distances can be mapped to probabilities using SoftMax function $p(d) = \frac{e^{-d}}{\sum_{d'}  e^{-d'}}$. 

\section{Training Setup}\label{sec:tasks}
ML models are data hungry. For code compliance assessment however, there are no large-scale datasets containing natural language policies paired with their compliant and non-compliant examples. We re-interpret and repurpose general datasets on code, documentation, code reviews, and bug-fixes with additional training schemes to bridge this gap. 

\subsection{Multi-Task Pre-training} 
To capture patterns in higher level documentation guidelines, pre-training with general documentation can be helpful. To capture coding issues and aligned code examples, pre-training with code reviews can be beneficial. We use below tasks.
\begin{itemize}
\item \textit{Collocated documentation prediction} (Doc): Paragraphs of software documentation are segmented into non-overlapping fixed-length passages. The passages belonging to the same paragraph are assigned a common label. The objective of  \textit{Collocated documentation prediction} is to predict which candidate segments (code snippet or text) in a batch come from the same paragraph. 
\item \textit{Code-comment matching task} (CC): Given multiple <code, review comment> pairs, we attempt to determine (a) the correct review comment given a code (\textit{relevant comment prediction}) and (b) the correct code on which a review comment is made (\textit{non-compliant code prediction}). 
\end{itemize}

\subsection{Pre-fine-tuning} 
Examples containing bug-fixes are especially valuable to us, because the example can be reinterpreted as a proxy to policy description with compliant and non-compliant codes. Such a dataset can then be used to pre-finetune for target task.  See Figure \ref{fig:cbf2} for a motivating example of a GitHub bug-fix review comment. The comment provides a security best practice suggestion for which the code-before and code-after versions can be used as non-compliant and compliant examples, respectively. Each review comment is essentially treated as a policy and the code-before and the code-after versions are taken to be the non-compliant and the compliant code examples respectively. Irrelevant examples  $c^\sim$ are mined from codes of unrelated bug-fixes. This dataset can be noisy, but we expect useful signals to emerge with large-scale datasets. 

\begin{figure*}
\caption*{Code review comment: \textit{Probably better to use `SecureRandom` for anything security related..} }
\begin{subfigure}[t]{0.48\textwidth}
\caption*{Facet: non-compliant/ Code-before}
\begin{lstlisting}[basicstyle=\scriptsize,language=Java,frame=single,breaklines=true,autogobble]
private String generatePassword(){
  String allowedChars="abcdefghijklmnopqrstuvwxyz0123456789";
  int charRange=allowedChars.length();
  Random random=new Random();
  StringBuilder password=new StringBuilder();
  int passwordLength=10;
  for (int i=0; i < passwordLength; i++) {
    password.append(allowedChars.charAt(random.nextInt(charRange)));
  }
  return password.toString();
}
\end{lstlisting}
\end{subfigure}
\hspace{0.4cm}
 \begin{subfigure}[t]{0.48\textwidth}
\caption*{Facet: compliant/ Code-after}
\begin{lstlisting}[basicstyle=\scriptsize,language=Java,frame=single,breaklines=true,autogobble]
private String generatePassword() throws NoSuchAlgorithmException {
  String allowedChars="abcdefghijklmnopqrstuvwxyz0123456789";
  int charRange=allowedChars.length();
  SecureRandom random=SecureRandom.getInstanceStrong();
  int passwordLength=20;
  char[] password=new char[passwordLength];
  for (int i=0; i < passwordLength; i++) {
    password[i]=allowedChars.charAt(random.nextInt(charRange));
  }
  return new String(password);
}
\end{lstlisting}
\end{subfigure}
\caption{Example entry in GitHub comment and bug-fix dataset}
\label{fig:cbf2}
\end{figure*}

\emph{Data Filtering}: As an additional treatment, we apply \emph{Doc2BP} \cite{doc2policy}, a deep learning classifier trained to detect coding policies and best practice recommendations from natural language text. \emph{Doc2BP} analyzes keywords, parts of speech patterns, and other linguistic properties to determine if a comment resembles coding policies, and filters out non-policy-like comments to reduce noise. Results with and without  \emph{Doc2BP} indicate that the filtering step in pre-finetuning is useful. 

\section{Evaluation Setup}\label{sec:benchmark}
We construct a new benchmark  by repurposing two listings of coding issues and a public code search dataset. 

\begin{itemize} 
\item Common Weaknesses Enumeration (\textit{CWE}) View 702 \cite{cweSite}  is a public domain list of software weaknesses introduced during implementation. It contains 185 Java weaknesses, each with a short natural language description and compliant and non-compliant examples. We use the weakness description as the policy. The number of policies with non-compliant, compliant, and both types of examples are 182, 36, and 33, respectively.  
\item Coding Best Practice (\textit{CBP}) is a collection of Java best practice descriptions and compliant and non-compliant examples obtained from Amazon CodeGuru \cite{codeguru} static analyzer test-suite. Our dataset contains 46 Java best practices, each with at least one compliant and one non-compliant example, and a total of 186 examples. 
\item CodeSearchNet (\textit{CSN}) is a public domain dataset \cite{husain2020codesearchnet}. We use the test split with 27K unlabeled code examples.  
\end{itemize} 

The resulting benchmark contains \textit{231 policies}, \textit{477 labeled codes}, and \textit{27K unlabeled codes} (total 28K code examples). Policies cover many topics like input validation, cipher security, web security, command injection, preferred data structures, etc. The dataset has no overlap with the training data, hence this is a zero-shot benchmark. We investigate two tasks: 
\begin{itemize} 
\item Compliance Classification: The objective is to predict whether a code is compliant, non-compliant, or irrelevant given a policy. Performance is measured in accuracy with respect to the known ground truth. 

\item Compliance Search: The objective is to retrieve compliant and non-compliant codes from a large code corpus for a user-specified natural language policy. In a large corpus setting, we do not expect to know the ground truth labels for all code examples. Hence we rely on another metric in information retrieval research, the mean reciprocal rank metric ($MRR$) defined as $\frac{1}{Q} \sum_{i=1}^{|Q|} \frac{1}{rank_i}$ where $Q$ denotes the number of user queries and $rank_i$ denotes the rank position of the first  example from the known ground truth within each facet. We prefer MRR over other information retrieval metrics such as precision@k for its ability to capture the order of relevance. For example, a precision@5 metric would yield the same score (0.2) whether a single relevant result appears at rank 1 or rank 5. MRR on the other hand, would generate a score of $1$ for the first position and a score of $0.2$ for the fifth position. 
\end{itemize} 

We propose two competitive baselines. 
\begin{itemize}
\item CodeBERT  \cite{feng2020codebert}  is the off-the-shelf pre-trained HuggingFace implementation. CodeBERT is a state-of-the-art deep learning approach for several code-related tasks including code search (\url{https://microsoft.github.io/CodeXGLUE/}). Despite using a token sequence representation of code, it is able to out-perform complex models such as GraphCodeBERT that can represent programs as graphs. This baseline is designed for relevance and not fine-grained differentiation. It represents the current best tool available to developers to find relevant examples, which can then be manually classified for compliance. 
\item Multi-class classification is a three-class (compliant, non-compliant, irrelevant) classifier. It is implemented by adding dropout and linear layers on top of the CodeBERT's \cite{feng2020codebert} transformer embedding layer. 
This is also trained for a zero-shot setting. A (policy, code) pair is represented using CodeBERT's input sequence convention 
$[CLS] x_1, x_2, ... x_m [SEP] y_1, y_2, ... y_n [SEP]$ where $x$ and $y$  represent the tokens for policy and code respectively.  
\end{itemize}

Our use of public domain datasets and easy to replicate baselines is expected to fuel  future research on this topic.

\section{Results}\label{sec:results}
This section presents a comprehensive evaluation. \textbf{Bold}-facing in tables is used to indicate superior performance. 

\subsection{Experimental Setup}
As the central bimodal code-text encoder for Policy2Code as well as the multi-class classification baseline, we use the HuggingFace implementation of CodeBERT \cite{feng2020codebert}. The multi-class classification baseline and the Policy2Code model are trained using the same pre-training and pre-fine-tuning data for apples-to-apples comparison. For the \textit{collocated documentation pre-training} (DOC) task, we use Java 8, 11, and 16 documentation ($\sim$1 GB). For \textit{code comment matching pre-training} (CC), we used 150K GitHub code reviews. For \textit{bug-fix-comment pre-fine-tuning}, we leveraged 32K code reviews on GitHub that were tagged as bug-fix, split into 80\%-20\% for training and validation. All these datasets  are in public domain. Further none of them intentionally overlap with benchmark examples. Code is statically represented as token sequences with sub-word tokenization and  BytePair Encoding (BPE) \cite{bpe}. 

\subsection{Policy2Code Model Training}
\begin{table*}
\begin{center}
 \caption{Effect of Loss Formulation and Policy Representation}
 \label{tab:lossAblation}
 \begin{tabular}{ | l  | l | c | c | c | c | c | c |}
\hline
 \multirow{2}{*}{Loss} &  \multirow{2}{*}{Policy Rep} &  \multicolumn{3}{|c|}{Classification Accuracy}   & \multicolumn{3}{|c|}{Search MRR}   \\
 & & CWE (Conceptual) & CWE & CBP & CWE (Conceptual)  & CWE & CBP     \\ \hline
 \multirow{2}{*}{QL}  & Prefixed        &  42.6 $\pm$ 3.6 & 48.13 $\pm$  2.2 & 44.93 $\pm$ 0.98  & 0.021 $\pm$ 0.00 & 0.0321 $\pm$ 0.009 & 0.0915 $\pm$ 0.017 \\
  & Masked          & 33.41 $\pm$ 1.9  &  34.97  $\pm$ 2.2 & 39.3 $\pm$ 6.3   & 0.0298 $\pm$ 0.02  &   0.0318 $\pm$ 0.19 & 0.07641 $\pm$ 0.029 \\ \hline
\multirow{2}{*}{BMT}  & Prefixed  &    \textbf{45.3} $\pm$ 2.4 & \textbf{49.8} $\pm$ 5.1 & \textbf{48.0} $\pm$ 7.1 &  0.0264 $\pm$ 0.029  &  \textbf{0.0363} $\pm$ 0.011 &  \textbf{0.1496} $\pm$ 0.021 \\ 
 & Masked          &  41.6 $\pm$ 4.5 & 42.23 $\pm$  3.7 & 46.21 $\pm$ 1.2 & 0.0242 $\pm$ 0.061  &  0.0213 $\pm$ 0.032 &  0.1127 $\pm$ 0.027  \\ \hline
\end{tabular}
\end{center}
\end{table*}
 
  \begin{table*}
\begin{center}
 \caption{Effect of Pre-training and Filtering Schemes}
 \label{tab:ablationTrain}
 \begin{tabular}{  | l | c | c | c | c | c | c |}
\hline
 \multirow{2}{*}{Pre-Training Scheme} &  \multicolumn{3}{|c|}{Classification Accuracy}   & \multicolumn{3}{|c|}{Search MRR}   \\
 & CWE (Conceptual) & CWE & CBP & CWE (Conceptual)  & CWE & CBP     \\ \hline
None    &  45.3 $\pm$ 2.4 & 49.8 $\pm$ 5.1 & 48.0 $\pm$ 7.1 & 0.0264 $\pm$ 0.029  & 0.0363 $\pm$ 0.011  & 0.1496 $\pm$ 0.021\\
Doc       &  48.3 $\pm$ 3.1 & 54.2 $\pm$ 4.6 & 56.2 $\pm$ 2.6 & 0.0319 $\pm$ 0.038  &  0.0313 $\pm$ 0.008  & 0.1773 $\pm$ 0.009 \\
CC     &  47.8 $\pm$ 6.1 & 43.77 $\pm$ 1.1 & 55.2 $\pm$ 2.9  & 0.0471 $\pm$ 0.129  &  0.0467 $\pm$ 0.008 & 0.1654 $\pm$ 0.017 \\
Doc + CC   &  53.2 $\pm$ 3.2  &  \textbf{59.3} $\pm$ 7.4  & 67.6 $\pm$ 7.9  & \textbf{0.0492} $\pm$ 0.05 &  0.0482 $\pm$ 0.011 &  0.1982 $\pm$ 0.012  \\
Doc + CC + \emph{Doc2BP} Filter \cite{doc2policy}  &   \textbf{53.8} $\pm$ 6.1  & \textbf{59.3} $\pm$ 2.0 & \textbf{71.0} $\pm$ 2.8 & 0.0462 ± 0.013  &  \textbf{0.0538} $\pm$ 0.01 &   \textbf{0.2182} $\pm$ 0.012   \\ \hline
\end{tabular}
\end{center}
\end{table*}
 
We now discuss the effect of policy representations, losses, and training schemes. Models were optimized using automatic hyper-parameter tuning. See Appendix \ref{sec:restuning} for details. 

Table \ref{tab:lossAblation} reports the performance of different combinations of loss functions and policy representations on the CWE and CBP datasets for classification and search tasks. \emph{The BMT loss with facet-prefixed policy representation is the most effective strategy across both datasets and tasks}.  
 
Table \ref{tab:ablationTrain} shows the effect of different pre-training and filtering schemes on Policy2Code performance. All schemes use BMT loss with facet-prefixed policy representation and subsequently followed by the same pre-finetuning step. The filtering step reduces the dataset to 14K examples (out of 32K original bug-fix dataset) which are predicted to be policy-like.  \emph{The pre-training schemes as well as the filtering step improve model performance across tasks}.  We continue to see better performance for the CBP dataset than the CWE dataset. We also observe that  documentation and code-related pre-training impact CWE performance differently. While code-related pre-training (CC)  decreases the CWE performance, documentation pre-training improves the performance. This effect can be explained as CWE policies are more conceptual, hence more similar to documentation patterns than specific code comments. 


\subsection{Baseline Comparison}
Table~\ref{tab:classification2} compares Policy2Code with baselines on compliance classification and compliance search tasks. 
Policy2Code outperforms CodeBERT as well as the multi-class classification baseline. Policy2Code achieves classification accuracies of (59\%, 71\%) and search MRR of (0.05, 0.21) on CWE and CBP.  On the same datasets, a popular CodeBERT baseline achieves classification accuracies of (37\%, 54\%) and MRR of (0.02, 0.02) respectively. Multi-class classification achieves accuracies of (54\%, 60\%) and MRR of (0.03, 0.13) respectively.  

We observe that the model performance is generally lower on the CWE dataset. Our investigation shows that CWE contains multiple policies on related topics and partially labeled ground truth. One code example may be applicable to multiple policies, but not all such relationships are recorded. For example, CWE-338\footnote{\url{https://cwe.mitre.org/data/definitions/338.html}} suggests \emph{using cryptographically strong Pseudo-Random Number Generator (PRNG) in a security context}. The below code is the only explicitly labeled example for this policy. 
\begin{lstlisting}[basicstyle=\scriptsize,language=Java,frame=single,breaklines=true,autogobble]
Random random = new Random(System.currentTimeMillis());
int accountID = random.nextInt();
\end{lstlisting}
However, other code examples in the CWE dataset may also be related to CWE-338. See the below code for example, labeled only for policy CWE-336\footnote{\url{https://cwe.mitre.org/data/definitions/336.html}} on \emph{not using the same seed across multiple PRNG initializations}. This example can also be considered as a valid example for CWE-338. 
\begin{lstlisting}[basicstyle=\scriptsize,language=Java,frame=single,breaklines=true,autogobble]
private static final long SEED = 1234567890;
public int generateAccountID() {
    Random random = new Random(SEED);
    return random.nextInt();
}
\end{lstlisting}
This implies that the CWE ground truth is incomplete, leading to currently computed metrics being a lower bound on the actual performance. CBP is expected to have higher quality as it is manually curated for compliance analysis task specifically. 

\begin{table*}
\begin{center}
 \caption{Comparison of Policy2Code with Baselines on Compliance Assessment Tasks}
 \label{tab:classification2}
 \begin{tabular}{  | l | c | c | c | c | c | c |}
\hline
 \multirow{2}{*}{Method} &  \multicolumn{3}{|c|}{Classification Accuracy}   & \multicolumn{3}{|c|}{Search MRR}   \\
 & CWE (Conceptual) & CWE & CBP & CWE (Conceptual)  & CWE & CBP     \\ \hline
CodeBERT   &  33.5  & 37.5 &  54.4 &  0.0223 & 0.0242 &  0.0217 \\  
Multi-class     & 42.82 $\pm$ 3.1  &  54.36  $\pm$ 1.9  & 60.04 $\pm$  2.7 &  0.0297 $\pm$ 0.03  & 0.0312 $\pm$ 0.02 & 0.1310 $\pm$ 0.03  \\ 
Policy2Code  & \textbf{53.84} $\pm$ 6.1  & \textbf{59.29} $\pm$ 2.0 & \textbf{71.04} $\pm$ 2.8 &   \textbf{0.0462} $\pm$ 0.013 & \textbf{0.0538} $\pm$ 0.01 &   \textbf{0.2182} $\pm$ 0.012 \\  \hline 
\end{tabular}
\end{center}
\end{table*}

\subsection{User Study}
We conducted a user study to determine if Policy2Code can be used in a practical setting for searching compliant and non-compliant code examples. One application for such examples is to serve as test cases for creating new static analysis rules.

We built a FAISS-GPU index \cite{faiss} over all 28K code examples in the benchmark dataset. On a p3.xlarge EC2 machine, index creation takes approximately 8 seconds. At inference time, top examples for any policy queries are retrieved within milliseconds based on approximate and efficient nearest neighbor algorithm. Exact similarity computation is avoided as it can take about 30 minutes per policy to compare all codes. 

We recruited 16 experts with an average of over 10 years of software engineering experience and over 5 years of programming language analysis experience. For 25 policies in CWE dataset, we determined the top 5 compliant and non-compliant examples detected by various approaches in the entire 28K code example benchmark dataset.  We masked the identity of the selection algorithm and asked the assessor if they agree or disagree with the label assignment. Table \ref{tab:shadowReview} shows the acceptance rate of predictions made by different techniques. \emph{Policy2Code has the highest overall  acceptance rate at 24\%, followed by 15.2\% for multi-class classification and 7.2\% for CodeBERT, averaged over all assessments.}

\begin{table}
\begin{center}
 \caption{Acceptance Rate in User Study (\%)}
 \label{tab:shadowReview}
 \begin{tabular}{|l|c|c|c|}
\hline
\textbf{Model} & \textbf{Compliant}  & \textbf{Non-compliant}  & \textbf{Overall} \\ \hline
CodeBERT  &  9.68  & 6.38	  &  7.2  \\
Multi-class & 27.01 & 11.57 & 15.2 \\
Policy2Code  & \textbf{41.93} & \textbf{17.64} & \textbf{24.13} \\  
\hline \end{tabular}
\end{center}
 \end{table}
 
\subsection{Anecdotal Results}
See Figures \ref{fig:moreConforming} and \ref{fig:moreViolating} for anecdotal examples. On the one hand, they may seem to over-index on keyword similarity (Figure \ref{fig:moreConforming}(c), (e)), and on the other hand, we find surprisingly semantic associations (Figure \ref{fig:moreConforming}(a)). Models trained on textual features alone, without using any advanced program representations may over-index on keyword similarity rather than higher order semantics. Here is an incorrect finding,  \textit{The software does not properly account for differences in case sensitivity when accessing or determining the properties of a resource, leading to inconsistent results.} (CWE-178).
\begin{lstlisting}[basicstyle=\scriptsize,language=Java,frame=single,breaklines=true,autogobble]
private boolean queryRelativeCatalogs () {
    if (resources==null) readProperties();
    if (resources==null) return defaultRelativeCatalogs;
    try {
        String allow = resources.getString(`catalogs'); 
        return (allow.equalsIgnoreCase(`true') || allow.equalsIgnoreCase(`y') || allow.equalsIgnoreCase(`1'));
    } catch (MissingResourceException e) {
        return defaultRelativeCatalogs;
    }
}
\end{lstlisting}
The repeated usage of keywords \textit{resource, properties, and case} in the result, leads to the example being scored higher, even though the notion of case-sensitivity is incorrectly modeled. Here, the case handling is a design choice. Policy2Code makes a mistake on this very challenging example.

\begin{figure}
\vspace{-0.2cm}
\caption*{\scriptsize a) \textit{Truncation errors occur when a primitive is cast to a primitive of a smaller size and data is lost in the conversion.} (CWE-197)}
\begin{lstlisting}[basicstyle=\scriptsize,language=Java,frame=single,breaklines=true,autogobble]
public static byte[] toByteArray(InputStream input, long size) throws IOException {
    if (size > Integer.MAX_VALUE) {
        throw new IllegalArgumentException(""Size greater than Integer max value: "" + size);
    }
    return toByteArray(input, (int) size);
}
\end{lstlisting}
\caption*{\scriptsize b) \textit{The software accepts XML from an untrusted source but does not validate the XML against the proper schema.} (CWE-112)}
\begin{lstlisting}[basicstyle=\scriptsize,language=Java,frame=single,breaklines=true,autogobble]
public static Document newDocument(String xmlString, boolean namespaceAware) throws SAXException, IOException, ParserConfigurationException {
    return XmlUtils.newDocument(new InputSource(new StringReader(xmlString)), namespaceAware);
}
\end{lstlisting}

\caption*{\scriptsize  c) \textit{This program compares classes by name, which can cause it to use the wrong class when multiple classes can have the same name.} (CWE-486)}
\begin{lstlisting}[basicstyle=\scriptsize,language=Java,frame=single,breaklines=true,autogobble]
protected static String createTableRefName(final Object entity) {
    Class type = entity.getClass();
    type = (type == Class.class ? (Class) entity : type);
    return (type.getSimpleName() + '_');
}
\end{lstlisting}

\caption*{\scriptsize  d) \textit{The product does not sufficiently enforce boundaries between the states of different sessions, causing data to be provided to, or used by, the wrong session.} (CWE-488)}
\begin{lstlisting}[basicstyle=\scriptsize,language=Java,frame=single,breaklines=true,autogobble]
protected void update(CollectionUpdateType force) throws IOException {
    State localState;
    synchronized (lock) {
        if (first) {
            state = checkState();
            state.last = System.currentTimeMillis();
            return;
        }
        localState = state.copy();
    }
    updateCollection(localState, force);
    localState.last = System.currentTimeMillis();
    synchronized (lock) {
        state = localState;
    }
}
\end{lstlisting}

\caption*{\scriptsize e) \textit{The product uses untrusted input when calculating or using an array index, but the product does not validate or incorrectly validates the index to ensure the index references a valid position within the array.} (CWE-129)}
\begin{lstlisting}[basicstyle=\scriptsize,language=Java,frame=single,breaklines=true,autogobble]
private static int calculateEndIndex( double[] array, int originalIndex ) {
    final int length = array.length;
    Exceptions.requireNonNull( array, ""array cannot be null"" );
    int index = originalIndex;
    if ( index < 0 ) {
        index = length + index;
    }
    if ( index < 0 ) {
        index = 0;
    }
    if ( index > length ) {
        index = length;
    }
    return index;
}
\end{lstlisting}

\caption{\textit{More Examples for \textbf{Compliant} Facet} - All examples are real detections from Policy2Code on CodeSearchNet-extended corpus. CWE ids are provided for reference only.}
\label{fig:moreConforming}
\end{figure}

\begin{figure}
\vspace{-0.1cm}
\caption*{\scriptsize a) \textit{Catching overly broad exceptions promotes complex error handling code that is more likely to contain security vulnerabilities.} (CWE-396)}
\begin{lstlisting}[basicstyle=\scriptsize,language=Java,frame=single,breaklines=true,autogobble]
private static byte[] decrypt(byte[] src, byte[] key) throws RuntimeException {
    try {
        SecureRandom sr = new SecureRandom();
        DESKeySpec dks = new DESKeySpec(key);
        SecretKeyFactory keyFactory = SecretKeyFactory.getInstance(DES);
        SecretKey securekey = keyFactory.generateSecret(dks);
        Cipher cipher = Cipher.getInstance(DES);
        cipher.init(Cipher.DECRYPT_MODE, securekey, sr);
        return cipher.doFinal(src);
    } catch (Exception e) {
        throw new RuntimeException(e);
    }
}
\end{lstlisting}

\caption*{\scriptsize b) \textit{Do not throw new checked exceptions (CheckedExceptions) forcing users to handle them elsewhere. Code should either propagate existing checked exceptions or handle them.}}
\begin{lstlisting}[basicstyle=\scriptsize,language=Java,frame=single,breaklines=true,autogobble]
public static Cipher newCipher(String algorithm) {
    try {
        return Cipher.getInstance(algorithm);
    }
    catch (NoSuchAlgorithmException e) {
        throw new IllegalArgumentException(""Not a valid encryption algorithm"", e);
    }
    catch (NoSuchPaddingException e) {
        throw new IllegalStateException(""Should not happen"", e);
    }
}
\end{lstlisting}

\caption*{\scriptsize c) \textit{Instead of repeatedly creating a new Random object to obtain multiple random numbers, create a single Random object and reuse it.}}
\begin{lstlisting}[basicstyle=\scriptsize,language=Java,frame=single,breaklines=true,autogobble]
private static final long SEED = 1234567890;
public int generateAccountID() {
    Random random = new Random(SEED);
    return random.nextInt();
}
\end{lstlisting}

\caption*{\scriptsize d) \textit{The product receives input that is expected to specify an index, position, or offset into an indexable resource such as a buffer or file, but it does not validate or incorrectly validates that the specified index/position/offset has the required properties.} (CWE-1285)}
\begin{lstlisting}[basicstyle=\scriptsize,language=Java,frame=single,breaklines=true,autogobble]
private static Object getCollectionProp(Object o, String propName, int index, String[] path) {
    o = _getFieldValuesFromCollectionOrArray(o, propName);
    if ( index + 1 == path.length ) {
        return o;
    } else {
        index++;
        return getCollectionProp(o, path[index], index, path);
    }
}
\end{lstlisting}

\caption*{\scriptsize e) \textit{The software does not properly handle when the expected number of parameters, fields,or arguments is not provided in input, or if those parameters are undefined.} (CWE-229)}
\begin{lstlisting}[basicstyle=\scriptsize,language=Java,frame=single,breaklines=true,autogobble]
public static <M extends Model> boolean isNew(M m, String pk_column) {
    final Object val = m.get(pk_column);
    return val == null || val instanceof Number && ((Number) val).intValue() <= 0;
}
\end{lstlisting}

\caption{\textit{More Examples for \textbf{Non-compliant} Facet} - All examples are real detections from Policy2Code on CodeSearchNet-extended corpus. CWE ids are provided for reference only.}
\label{fig:moreViolating}
\end{figure}

\section{Threats to Validity}\label{sec:threats}
As we formulate code compliance assessment as a machine learning problem, we see certain threats to the validity. 
\begin{itemize}
\item ML models are known to carry biases from the training datasets. While more training schemes (pre-training as well as fine-tuning) can help improve model performance, some policies may still be under-represented in these training datasets. Certain issues may rarely occur in code and even rarely detected by human reviewers. The models trained on human code review comments and bug-fixes may not learn to detect these issues well. 
\item The benchmark dataset needs further curation. In using the CWE dataset, we see that certain policy descriptions are conceptual and loosely scripted that do not convey a strong recommendation or warning. Further, some ground truth labels are missing. Future work should further enhance the benchmark in both aspects to improve assessment. 
\item Finally, ML predictions may never meet 100\% accuracy unlike carefully crafted static analysis rules. We expect automated tools to augment and not replace manual code reviews and static analyzer creation. Some reduced degree of manual vetting may still be required. 
\end{itemize}  

We need to understand how the performance of machine learning algorithms vary by data characteristics (policy and code) and how it can be improved by innovating on natural language (NL) and programming language (PL) research.

\section{Conclusion}\label{sec:conclusion} 
Code compliance assessment  is an important problem in software development and an emerging area for machine learning. This paper explores novel research questions related to learning framework, training data, and evaluation setup. We proposed a representation learning approach that preserves the relationships between policies and their compliant, non-compliant, and irrelevant code examples via the vector distances between their embeddings. To overcome the lack of task-specific training data, we proposed repurposing general software datasets with pre-training, pre-fine-tuning, and filtering steps. We evaluated the impact of policy representations, losses, training schemes, and hyper-parameter optimization. Resulting Policy2Code model shows promising results for compliance classification and search tasks. Policy2Code achieves classification accuracies of (59\%, 71\%) and search MRR of (0.05, 0.21) on CWE and CBP, the two types of datasets in our benchmark.  On the same datasets,  CodeBERT baseline achieves classification accuracies of (37\%, 54\%) and MRR of (0.02, 0.02) respectively whereas multi-class  baseline achieves classification accuracies of (54\%, 60\%) and MRR of (0.03, 0.13) respectively.  In a user study of compliant and non-compliant findings, 24\% detections from Policy2Code were accepted compared to only 7\% detections from CodeBERT and 15\% by the multi-class  baseline. 

We hope to encourage more theoretical and empirical research in automated code compliance assessment. We expect ML solutions to significantly reduce future manual efforts in code reviews and static analysis. Improvements can come from innovating on input transformations, use of program analysis and domain knowledge (API knowledge graph), and robust ML formulations. 

\appendices 

\section{Facet-masked Policy Representation}\label{sec:facetLearn}
The idea for facet-masked representation is based on the conditional masking work, originally used in computer vision to compute image similarities across aspects such as color and texture~ \cite{veit2017csn}. The embedding of policy $r$ is factorized depending on facet value $y$. The factorization is implemented via a  conditional mask $\textbf{m} \in R^{d \times n_k}$, where $d$ is the embedding vector length and $n_k$ is the number of facets (=2). The mask can be paramterized as $\textbf{m} = \sigma(\beta)$, with $\sigma$ denoting a rectified linear unit so that $\sigma(\beta) = max\{0, \beta\}$. The $k^{th}$ column $m_k$ plays the role of an element-wise gating function selecting the relevant subspace of the embedding dimensions to attend to the $k^{th}$ facet.  Additional loss regularization terms are used  to encourage embeddings to be drawn from a unit ball ($L_W$) and for mask regularization ($L_M$).

\section{Model Tuning Experiments}\label{sec:restuning}

Performance of neural models can be significantly improved with proper hyper-parameter tuning (HPT). We used automatic selection of parameters (batch size, learning rate, etc) based on a validation dataset and experimented with triplet loss variations.  
\begin{itemize} 
\item \textit{Batch All Triplet} loss computes the loss for all possible, valid  triplets in the batch, i.e., anchor and positive must have the same label, anchor and negative a different label. 
\item \textit{Batch Hard Triplet} loss computes the loss for all possible, valid triplets. It then looks for the hardest positive (largest $d(f_a, f_p)$) and the hardest negatives (smallest $d(f_a, f_n)$) per label class, and sums the loss only over them. 
\item \textit{Batch Semi-hard Triplet} loss computes the loss for all possible, valid  triplets. It then looks for the semi hard positives and negatives and sums the loss only over them. 
\item \textit{Batch Hard Triplet Soft-Margin} loss is a variation of the \textit{Batch Hard} triplet loss where the loss over the hardest positive and the hardest negative examples are computed with a soft margin, e.g. $\log1p(\exp(d(f_a, f_p) - d(f_a, f_n)))$. 
\end{itemize}

The determination of hard, semi-hard, and easy triplets was made with a tunable margin parameter. 
\begin{itemize}
\item Easy:  $d(f_a - f_p) + \alpha  < d (f_a - f_n) $. These are already well separated and not useful for model training.
\item Medium:  $d(f_a - f_p) < d(f_a - f_n) < d(f_a - f_p) + \alpha$. These triplets are typically more suited for optimisation.
\item Hard:  $d(f_a - f_n) < d(f_a - f_p)$. Selecting the hardest negative examples can sometimes lead to bad local minima and result in a collapsed model training \cite{Manmatha2017SamplingMI}.
\end{itemize}

\begin{figure}

\begin{minipage}{\columnwidth}
\centering
         \begin{minipage}[t]{0.7\columnwidth}
    	\centering
   	\includegraphics[width=\columnwidth]{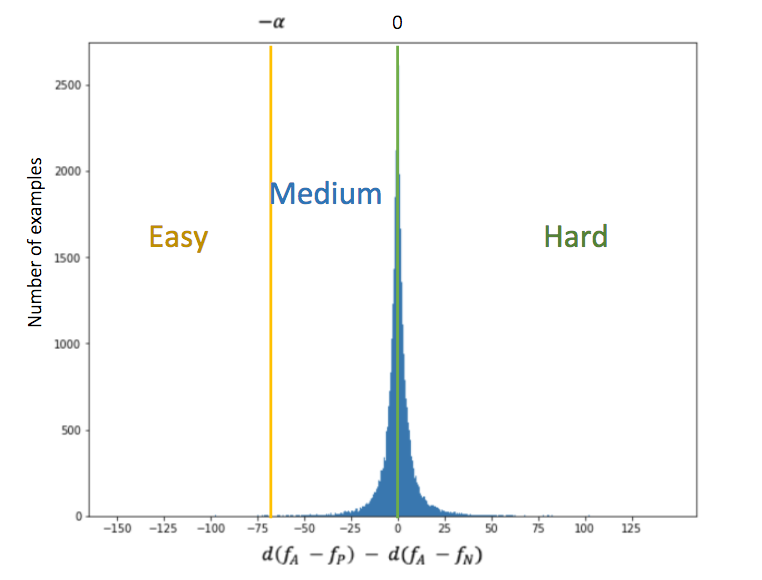}
       \captionof{figure}{Distribution of triplet distances can be used to tune margins and mine triplets.   \label{fig:bugFixDistro1}}
 	\end{minipage}
\quad \\ \vspace{0.2cm}
      \begin{minipage}[t]{0.7\columnwidth}
    \centering
      \includegraphics[width=\columnwidth]{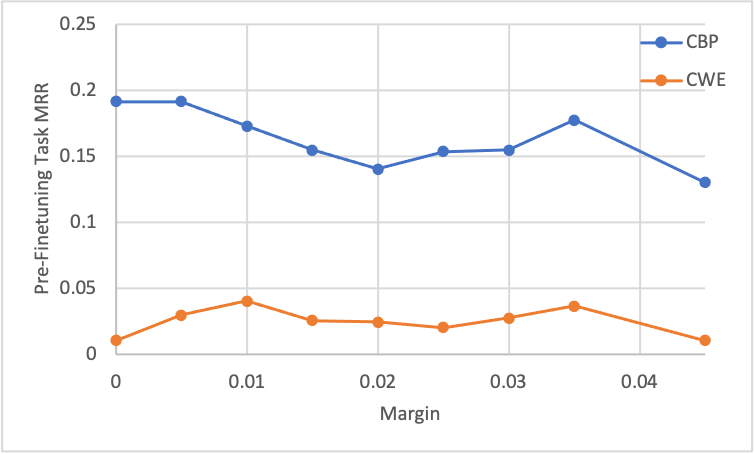}     
    \captionof{figure}{Impact of margin on the pre-fine-tuning MRR of \textit{Batch All Triplet} loss. \label{fig:margin1}}
  \end{minipage}
\end{minipage}

\end{figure}

 Figure \ref{fig:bugFixDistro1} shows the distribution of distances between the embeddings of text anchors and positive and negative code example for a sample of the dataset. The distribution is partitioned in three sections based on the margin, used to define easy, medium, and hard triplets. Figure \ref{fig:margin1} shows the effect of margin with various triplet modelling strategies. Table~\ref{tab:triplet} compares the performance of loss variations on fine-tuning the CodeBERT model. The margin parameter is tuned for each variation except the soft-margin formulation which does not use a margin threshold. We observe that \textit{Batch All Triplet loss has the best performance} and use it for all subsequent experiments. 
 
\begin{table}
 \caption{Performance of triplet loss variations}
 \label{tab:triplet}

  \begin{tabular}[t]{|l|c|c|}
\hline
Batch Triplet Loss Type & Overall Accuracy & Overall MRR \\ 
\hline
 All Triplets & \textbf{57.2 $\pm$ 6.2} &  \textbf{0.080  $\pm$ 0.02} \\ 
 Hard Triplets & 48.1 $\pm$ 4.7 & 0.063  $\pm$ 0.01\\
 Semi-Hard Triplets & 50.13 $\pm$ 2.8 & 0.068  $\pm$ 0.02\\
 Hard Triplets Soft-Margin & 53.12 $\pm$ 5.1 & 0.065  $\pm$ 0.04 \\ 
\hline \end{tabular}

 \end{table}

\bibliographystyle{IEEEtran}

\begin{thebibliography}{10}
\providecommand{\url}[1]{#1}
\csname url@samestyle\endcsname
\providecommand{\newblock}{\relax}
\providecommand{\bibinfo}[2]{#2}
\providecommand{\BIBentrySTDinterwordspacing}{\spaceskip=0pt\relax}
\providecommand{\BIBentryALTinterwordstretchfactor}{4}
\providecommand{\BIBentryALTinterwordspacing}{\spaceskip=\fontdimen2\font plus
\BIBentryALTinterwordstretchfactor\fontdimen3\font minus
  \fontdimen4\font\relax}
\providecommand{\BIBforeignlanguage}[2]{{%
\expandafter\ifx\csname l@#1\endcsname\relax
\typeout{** WARNING: IEEEtran.bst: No hyphenation pattern has been}%
\typeout{** loaded for the language `#1'. Using the pattern for}%
\typeout{** the default language instead.}%
\else
\language=\csname l@#1\endcsname
\fi
#2}}
\providecommand{\BIBdecl}{\relax}
\BIBdecl

\bibitem{cweSite}
{Mitre Corporation}, ``{The Common Weakness Enumeration (CWE) Initiative},''
  \url{http://cwe.mitre.org/}, 2021, [Online; accessed 2021].

\bibitem{marketResearch}
{Industry Research}, ``Global static code analysis software market report,
  history and forecast 2016-2027, breakdown data by companies, key regions,
  types and application,''
  \url{https://www.industryresearch.biz/global-static-code-analysis-software-market-18726250},
  p. 105, 2021, published: 2021-07-12.

\bibitem{habib2019neural}
\BIBentryALTinterwordspacing
A.~Habib and M.~Pradel, ``Neural bug finding: {A} study of opportunities and
  challenges,'' \emph{CoRR}, vol. abs/1906.00307, 2019. [Online]. Available:
  \url{http://arxiv.org/abs/1906.00307}
\BIBentrySTDinterwordspacing

\bibitem{pradel2018deepbugs}
\BIBentryALTinterwordspacing
M.~Pradel and K.~Sen, ``Deepbugs: A learning approach to name-based bug
  detection,'' \emph{Proc. ACM Program. Lang.}, vol.~2, no. OOPSLA, Oct. 2018.
  [Online]. Available: \url{https://doi.org/10.1145/3276517}
\BIBentrySTDinterwordspacing

\bibitem{neuralBugWithContextCodeRep}
Y.~Li, S.~Wang, T.~N. Nguyen, and S.~Van~Nguyen, ``Improving bug detection via
  context-based code representation learning and attention-based neural
  networks,'' \emph{Proc. ACM Program. Lang.}, vol.~3, no. OOPSLA, Oct. 2019.

\bibitem{9252010}
N.~Stulova, A.~Blasi, A.~Gorla, and O.~Nierstrasz, ``Towards detecting
  inconsistent comments in java source code automatically,'' in \emph{2020 IEEE
  20th International Working Conference on Source Code Analysis and
  Manipulation (SCAM)}, 2020, pp. 65--69.

\bibitem{sachdev2018ncs}
\BIBentryALTinterwordspacing
S.~Sachdev, H.~Li, S.~Luan, S.~Kim, K.~Sen, and S.~Chandra, ``Retrieval on
  source code: A neural code search,'' in \emph{Proceedings of the 2nd ACM
  SIGPLAN International Workshop on Machine Learning and Programming
  Languages}, ser. MAPL 2018.\hskip 1em plus 0.5em minus 0.4em\relax New York,
  NY, USA: Association for Computing Machinery, 2018, p. 31–41. [Online].
  Available: \url{https://doi.org/10.1145/3211346.3211353}
\BIBentrySTDinterwordspacing

\bibitem{cambronero2019WDLMCSmisc}
J.~Cambronero, H.~Li, S.~Kim, K.~Sen, and S.~Chandra, ``When deep learning met
  code search,'' 2019.

\bibitem{gu2018DCS}
\BIBentryALTinterwordspacing
X.~Gu, H.~Zhang, and S.~Kim, ``Deep code search,'' in \emph{Proceedings of the
  40th International Conference on Software Engineering}, ser. ICSE '18.\hskip
  1em plus 0.5em minus 0.4em\relax New York, NY, USA: Association for Computing
  Machinery, 2018, p. 933–944. [Online]. Available:
  \url{https://doi.org/10.1145/3180155.3180167}
\BIBentrySTDinterwordspacing

\bibitem{zeroshot}
H.~Larochelle, D.~Erhan, and Y.~Bengio, ``Zero-data learning of new tasks,'' in
  \emph{Proceedings of the 23rd National Conference on Artificial Intelligence
  - Volume 2}, ser. AAAI'08.\hskip 1em plus 0.5em minus 0.4em\relax Chicago,
  Illinois: AAAI Press, 2008, p. 646–651.

\bibitem{codeguru}
{Amazon Web Services}, \emph{Amazon CodeGuru: Automate code reviews and optimize
  application performance with ML-powered recommendations}, 2021,
  \url{https://aws.amazon.com/codeguru/}.

\bibitem{husain2020codesearchnet}
\BIBentryALTinterwordspacing
H.~Husain, H.~Wu, T.~Gazit, M.~Allamanis, and M.~Brockschmidt, ``Codesearchnet
  challenge: Evaluating the state of semantic code search,'' \emph{CoRR}, vol.
  abs/1909.09436, 2019. [Online]. Available:
  \url{http://arxiv.org/abs/1909.09436}
\BIBentrySTDinterwordspacing

\bibitem{allamanis2018learning}
\BIBentryALTinterwordspacing
M.~Allamanis, M.~Brockschmidt, and M.~Khademi, ``Learning to represent programs
  with graphs,'' in \emph{International Conference on Learning
  Representations}.\hskip 1em plus 0.5em minus 0.4em\relax Vancouver, BC,
  Canada: OpenReview.net, 2018. [Online]. Available:
  \url{https://openreview.net/forum?id=BJOFETxR-}
\BIBentrySTDinterwordspacing

\bibitem{9610635}
P.~Rani, M.~Birrer, S.~Panichella, M.~Ghafari, and O.~Nierstrasz, ``What do
  developers discuss about code comments?'' in \emph{2021 IEEE 21st
  International Working Conference on Source Code Analysis and Manipulation
  (SCAM)}, 2021, pp. 153--164.

\bibitem{siow2019core}
\BIBentryALTinterwordspacing
J.~K. Siow, C.~Gao, L.~Fan, S.~Chen, and Y.~Liu, ``{CORE:} automating review
  recommendation for code changes,'' \emph{CoRR}, vol. abs/1912.09652, 2019.
  [Online]. Available: \url{http://arxiv.org/abs/1912.09652}
\BIBentrySTDinterwordspacing

\bibitem{dlForCodeModelGen}
\BIBentryALTinterwordspacing
T.~H.~M. Le, H.~Chen, and M.~A. Babar, ``Deep learning for source code modeling
  and generation: Models, applications, and challenges,'' \emph{ACM Comput.
  Surv.}, vol.~53, no.~3, Jun. 2020. [Online]. Available:
  \url{https://doi.org/10.1145/3383458}
\BIBentrySTDinterwordspacing

\bibitem{9610724}
A.~Naghshzan, L.~Guerrouj, and O.~Baysal, ``Leveraging unsupervised learning to
  summarize apis discussed in stack overflow,'' in \emph{2021 IEEE 21st
  International Working Conference on Source Code Analysis and Manipulation
  (SCAM)}, 2021, pp. 142--152.

\bibitem{attentionIsAllYouNeed}
A.~Vaswani, N.~Shazeer, N.~Parmar, J.~Uszkoreit, L.~Jones, A.~N. Gomez,
  u.~Kaiser, and I.~Polosukhin, ``Attention is all you need,'' in
  \emph{Proceedings of the 31st International Conference on Neural Information
  Processing Systems}, ser. NIPS'17.\hskip 1em plus 0.5em minus 0.4em\relax Red
  Hook, NY, USA: Curran Associates Inc., 2017, p. 6000–6010.

\bibitem{feng2020codebert}
\BIBentryALTinterwordspacing
Z.~Feng, D.~Guo, D.~Tang, N.~Duan, X.~Feng, M.~Gong, L.~Shou, B.~Qin, T.~Liu,
  D.~Jiang, and M.~Zhou, ``{C}ode{BERT}: A pre-trained model for programming
  and natural languages,'' in \emph{Findings of the Association for
  Computational Linguistics: EMNLP 2020}.\hskip 1em plus 0.5em minus
  0.4em\relax Online: Association for Computational Linguistics, Nov. 2020, pp.
  1536--1547. [Online]. Available:
  \url{https://www.aclweb.org/anthology/2020.findings-emnlp.139}
\BIBentrySTDinterwordspacing

\bibitem{devlin-etal-2019-bert}
\BIBentryALTinterwordspacing
J.~Devlin, M.-W. Chang, K.~Lee, and K.~Toutanova, ``{BERT}: Pre-training of
  deep bidirectional transformers for language understanding,'' in
  \emph{Proceedings of the 2019 Conference of the North {A}merican Chapter of
  the Association for Computational Linguistics: Human Language Technologies,
  Volume 1 (Long and Short Papers)}.\hskip 1em plus 0.5em minus 0.4em\relax
  Minneapolis, Minnesota: Association for Computational Linguistics, Jun. 2019,
  pp. 4171--4186. [Online]. Available:
  \url{https://www.aclweb.org/anthology/N19-1423}
\BIBentrySTDinterwordspacing

\bibitem{ramachandran-etal-2017-unsupervised}
\BIBentryALTinterwordspacing
P.~Ramachandran, P.~Liu, and Q.~Le, ``Unsupervised pretraining for sequence to
  sequence learning,'' in \emph{Proceedings of the 2017 Conference on Empirical
  Methods in Natural Language Processing}.\hskip 1em plus 0.5em minus
  0.4em\relax Copenhagen, Denmark: Association for Computational Linguistics,
  Sep. 2017, pp. 383--391. [Online]. Available:
  \url{https://www.aclweb.org/anthology/D17-1039}
\BIBentrySTDinterwordspacing

\bibitem{muppet}
A.~Aghajanyan, A.~Gupta, A.~Shrivastava, X.~Chen, L.~Zettlemoyer, and S.~Gupta,
  ``Muppet: Massive multi-task representations with pre-finetuning,'' 2021.

\bibitem{linstead2009Sourcerer}
\BIBentryALTinterwordspacing
E.~Linstead, S.~Bajracharya, T.~Ngo, P.~Rigor, C.~Lopes, and P.~Baldi,
  ``Sourcerer: Mining and searching internet-scale software repositories,''
  \emph{Data Min. Knowl. Discov.}, vol.~18, no.~2, p. 300–336, Apr. 2009.
  [Online]. Available: \url{https://doi.org/10.1007/s10618-008-0118-x}
\BIBentrySTDinterwordspacing

\bibitem{lv2015codehow}
\BIBentryALTinterwordspacing
F.~Lv, H.~Zhang, J.-g. Lou, S.~Wang, D.~Zhang, and J.~Zhao, ``Codehow:
  Effective code search based on api understanding and extended boolean
  model,'' in \emph{Proceedings of the 30th IEEE/ACM International Conference
  on Automated Software Engineering}, ser. ASE '15.\hskip 1em plus 0.5em minus
  0.4em\relax Lincoln, Nebraska: IEEE Press, 2015, p. 260–270. [Online].
  Available: \url{https://doi.org/10.1109/ASE.2015.42}
\BIBentrySTDinterwordspacing

\bibitem{mcmillan2011portfolio}
\BIBentryALTinterwordspacing
C.~Mcmillan, D.~Poshyvanyk, M.~Grechanik, Q.~Xie, and C.~Fu, ``Portfolio:
  Searching for relevant functions and their usages in millions of lines of
  code,'' \emph{ACM Trans. Softw. Eng. Methodol.}, vol.~22, no.~4, oct 2013.
  [Online]. Available: \url{https://doi.org/10.1145/2522920.2522930}
\BIBentrySTDinterwordspacing

\bibitem{raghothaman2016swim}
M.~{Raghothaman}, Y.~{Wei}, and Y.~{Hamadi}, ``Swim: Synthesizing what i mean -
  code search and idiomatic snippet synthesis,'' in \emph{2016 IEEE/ACM 38th
  International Conference on Software Engineering (ICSE)}.\hskip 1em plus
  0.5em minus 0.4em\relax Austin, TX, USA: IEEE, 2016, pp. 357--367.

\bibitem{lu2015codeWordNet}
M.~{Lu}, X.~{Sun}, S.~{Wang}, D.~{Lo}, and {Yucong Duan}, ``Query expansion via
  wordnet for effective code search,'' in \emph{2015 IEEE 22nd International
  Conference on Software Analysis, Evolution, and Reengineering (SANER)},
  no.~22.\hskip 1em plus 0.5em minus 0.4em\relax Montreal, CA: IEEE, 2015, pp.
  545--549.

\bibitem{allamanis2015bimodal}
M.~Allamanis, D.~Tarlow, A.~D. Gordon, and Y.~Wei, ``Bimodal modelling of
  source code and natural language,'' in \emph{Proceedings of the 32nd
  International Conference on International Conference on Machine Learning -
  Volume 37}, ser. ICML'15.\hskip 1em plus 0.5em minus 0.4em\relax Lille,
  France: JMLR.org, 2015, p. 2123–2132.

\bibitem{yao2019mman}
\BIBentryALTinterwordspacing
Y.~Wan, J.~Shu, Y.~Sui, G.~Xu, Z.~Zhao, J.~Wu, and P.~S. Yu, ``Multi-modal
  attention network learning for semantic source code retrieval,'' in
  \emph{Proceedings of the 34th IEEE/ACM International Conference on Automated
  Software Engineering}, ser. ASE '19, no.~34.\hskip 1em plus 0.5em minus
  0.4em\relax San Diego, California: IEEE Press, 2019, p. 13–25. [Online].
  Available: \url{https://doi.org/10.1109/ASE.2019.00012}
\BIBentrySTDinterwordspacing

\bibitem{ling2020deep}
\BIBentryALTinterwordspacing
X.~Ling, L.~Wu, S.~Wang, G.~Pan, T.~Ma, F.~Xu, A.~X. Liu, C.~Wu, and S.~Ji,
  ``Deep graph matching and searching for semantic code retrieval,''
  \emph{CoRR}, vol. abs/2010.12908, 2020. [Online]. Available:
  \url{https://arxiv.org/abs/2010.12908}
\BIBentrySTDinterwordspacing

\bibitem{9252051}
J.~P. Diniz, D.~Cruz, F.~Ferreira, C.~Tavares, and E.~Figueiredo, ``Github
  label embeddings,'' in \emph{2020 IEEE 20th International Working Conference
  on Source Code Analysis and Manipulation (SCAM)}, 2020, pp. 249--253.

\bibitem{cambronero2019WDLMCS}
\BIBentryALTinterwordspacing
J.~Cambronero, H.~Li, S.~Kim, K.~Sen, and S.~Chandra, ``When deep learning met
  code search,'' in \emph{Proceedings of the 2019 27th ACM Joint Meeting on
  European Software Engineering Conference and Symposium on the Foundations of
  Software Engineering}, ser. ESEC/FSE 2019.\hskip 1em plus 0.5em minus
  0.4em\relax New York, NY, USA: Association for Computing Machinery, 2019, p.
  964–974. [Online]. Available: \url{https://doi.org/10.1145/3338906.3340458}
\BIBentrySTDinterwordspacing

\bibitem{martins2020concra}
\BIBentryALTinterwordspacing
M.~de~Rezende~Martins and M.~A. Gerosa, ``Concra: A convolutional neural
  networks code retrieval approach,'' in \emph{Proceedings of the 34th
  Brazilian Symposium on Software Engineering}, ser. SBES '20.\hskip 1em plus
  0.5em minus 0.4em\relax New York, NY, USA: Association for Computing
  Machinery, 2020, p. 526–531. [Online]. Available:
  \url{https://doi.org/10.1145/3422392.3422462}
\BIBentrySTDinterwordspacing

\bibitem{yao2019coacor}
\BIBentryALTinterwordspacing
Z.~Yao, J.~R. Peddamail, and H.~Sun, ``Coacor: Code annotation for code
  retrieval with reinforcement learning,'' in \emph{The World Wide Web
  Conference}, ser. WWW '19.\hskip 1em plus 0.5em minus 0.4em\relax New York,
  NY, USA: Association for Computing Machinery, 2019, p. 2203–2214. [Online].
  Available: \url{https://doi.org/10.1145/3308558.3313632}
\BIBentrySTDinterwordspacing

\bibitem{Kaya2019DeepML}
M.~Kaya and H.~Bilge, ``Deep metric learning: A survey,'' \emph{Symmetry},
  vol.~11, p. 1066, 2019.

\bibitem{luan2020sparse}
\BIBentryALTinterwordspacing
Y.~Luan, J.~Eisenstein, K.~Toutanova, and M.~Collins, ``{Sparse, Dense, and
  Attentional Representations for Text Retrieval},'' \emph{Transactions of the
  Association for Computational Linguistics}, vol.~9, pp. 329--345, 04 2021.
  [Online]. Available: \url{https://doi.org/10.1162/tacl\_a\_00369}
\BIBentrySTDinterwordspacing

\bibitem{facetedSearch}
D.~Tunkelang, ``Faceted search,'' in \emph{Faceted Search}.\hskip 1em plus
  0.5em minus 0.4em\relax NY, USA: Morgan and Claypool, 2009.

\bibitem{beyondFaceted}
O.~Ben~Yitzhak, N.~Golbandi, N.~Har'El, R.~Lempel, A.~Neumann, S.~Ofek-Koifman,
  D.~Sheinwald, E.~Shekita, B.~Sznajder, and S.~Yogev, ``Beyond basic faceted
  search,'' \emph{WSDM'08 - Proceedings of the 2008 International Conference on
  Web Search and Data Mining}, vol.~8, pp. 33--44, 01 2008.

\bibitem{Manmatha2017SamplingMI}
R.~Manmatha, C.-Y. Wu, A.~Smola, and P.~Kr{\"a}henb{\"u}hl, ``Sampling matters
  in deep embedding learning,'' \emph{2017 IEEE International Conference on
  Computer Vision (ICCV)}, vol.~1, no.~1, pp. 2859--2867, 2017.

\bibitem{SchroffKP15}
\BIBentryALTinterwordspacing
F.~Schroff, D.~Kalenichenko, and J.~Philbin, ``Facenet: {A} unified embedding
  for face recognition and clustering,'' \emph{CoRR}, vol. abs/1503.03832,
  2015. [Online]. Available: \url{http://arxiv.org/abs/1503.03832}
\BIBentrySTDinterwordspacing

\bibitem{hermans2017defense}
A.~{Hermans}, L.~{Beyer}, and B.~{Leibe}, ``{In Defense of the Triplet Loss for
  Person Re-Identification},'' \emph{arXiv e-prints}, vol.~1, no.~1, Mar. 2017.

\bibitem{chen2017quadruplet}
\BIBentryALTinterwordspacing
W.~Chen, X.~Chen, J.~Zhang, and K.~Huang, ``Beyond triplet loss: a deep
  quadruplet network for person re-identification,'' \emph{CoRR}, vol.
  abs/1704.01719, 2017. [Online]. Available:
  \url{http://arxiv.org/abs/1704.01719}
\BIBentrySTDinterwordspacing

\bibitem{luo-etal-2019-towards}
\BIBentryALTinterwordspacing
F.~Luo, P.~Li, P.~Yang, J.~Zhou, Y.~Tan, B.~Chang, Z.~Sui, and X.~Sun,
  ``Towards fine-grained text sentiment transfer,'' in \emph{Proceedings of the
  57th Annual Meeting of the Association for Computational Linguistics}.\hskip
  1em plus 0.5em minus 0.4em\relax Florence, Italy: Association for
  Computational Linguistics, Jul. 2019, pp. 2013--2022. [Online]. Available:
  \url{https://aclanthology.org/P19-1194}
\BIBentrySTDinterwordspacing

\bibitem{DBLP:journals/corr/abs-1804-06437}
\BIBentryALTinterwordspacing
J.~Li, R.~Jia, H.~He, and P.~Liang, ``Delete, retrieve, generate: {A} simple
  approach to sentiment and style transfer,'' \emph{CoRR}, vol. abs/1804.06437,
  2018. [Online]. Available: \url{http://arxiv.org/abs/1804.06437}
\BIBentrySTDinterwordspacing

\bibitem{raffel2020T5}
\BIBentryALTinterwordspacing
C.~Raffel, N.~Shazeer, A.~Roberts, K.~Lee, S.~Narang, M.~Matena, Y.~Zhou,
  W.~Li, and P.~J. Liu, ``Exploring the limits of transfer learning with a
  unified text-to-text transformer,'' \emph{Journal of Machine Learning
  Research}, vol.~21, no. 140, pp. 1--67, 2020. [Online]. Available:
  \url{http://jmlr.org/papers/v21/20-074.html}
\BIBentrySTDinterwordspacing

\bibitem{veit2017csn}
A.~{Veit}, S.~{Belongie}, and T.~{Karaletsos}, ``Conditional similarity
  networks,'' in \emph{2017 IEEE Conference on Computer Vision and Pattern
  Recognition (CVPR)}, vol.~1.\hskip 1em plus 0.5em minus 0.4em\relax HI, USA:
  IEEE, 2017, pp. 1781--1789.

\bibitem{doc2policy}
N.~Sawant and S.~H. Sengamedu, ``Learning-based identification of coding best
  practices from software documentation,'' in \emph{2022 IEEE International
  Conference on Software Maintenance and Evolution}.\hskip 1em plus 0.5em minus
  0.4em\relax Limassol, Cyprus: IEEE Computer Society, oct 2022, pp.~--.

\bibitem{bpe}
\BIBentryALTinterwordspacing
R.~Sennrich, B.~Haddow, and A.~Birch, ``Neural machine translation of rare
  words with subword units,'' in \emph{Proceedings of the 54th Annual Meeting
  of the Association for Computational Linguistics (Volume 1: Long
  Papers)}.\hskip 1em plus 0.5em minus 0.4em\relax Berlin, Germany: Association
  for Computational Linguistics, Aug. 2016, pp. 1715--1725. [Online].
  Available: \url{https://www.aclweb.org/anthology/P16-1162}
\BIBentrySTDinterwordspacing

\bibitem{faiss}
J.~Johnson, M.~Douze, and H.~J{\'e}gou, ``Billion-scale similarity search with
  gpus,'' \emph{IEEE Transactions on Big Data}, vol.~-, no.~-, pp. 1--1, 2019.

\end{thebibliography}


\end{document}